\documentclass[aps,showpacs,superscriptaddress,groupedaddress,amsmath,amssymb,11pt]{revtex4}  
\usepackage{graphicx}  
\usepackage{dcolumn}   
\usepackage{bm}        
\usepackage{amssymb}   

\newcommand{\be}{\begin{equation}}
\newcommand{\en}{\end{equation}}
\newcommand{\beq}{\begin{eqnarray}}
\newcommand{\eeq}{\end{eqnarray}}
\newcommand{\ds}{\displaystyle}

\begin{document}

\title{Semi-holographic model including the radiation component}
\author{Sergio del Campo}
\email{sdelcamp@ucv.cl}
\affiliation{ Instituto de F\'{\i}sica,
Pontificia Universidad Cat\'olica de Valpara\'iso, Av. Universidad 330,
Curauma, Valpara\'iso, Chile,}

\author{V\'ictor H. C\'ardenas}
\email{victor.cardenas@uv.cl}
\affiliation{ Instituto de F\'{\i}sica y Astronom\'ia,
Universidad de Valpara\'iso, Gran Breta\~na 1111,
Valpara\'iso, Chile,}
\affiliation{Centro de Astrof\'isica de Valpara\'iso, Gran Breta\~na 1111, Playa Ancha,
\\ Valpara\'{\i}so, Chile.}

\author{Juan Maga\~na}%
\email{juan.magana@uv.cl}
\affiliation{ Instituto de F\'{\i}sica y Astronom\'ia,  Universidad de Valpara\'iso, Gran Breta\~na 1111,
Valpara\'iso, Chile,}
\affiliation{Centro de Astrof\'isica de Valpara\'iso, Gran Breta\~na 1111, Playa Ancha,
\\ Valpara\'{\i}so, Chile.}

\author{J. R. Villanueva}
\email{jose.villanuevalob@uv.cl}
\affiliation{ Instituto de F\'{\i}sica y Astronom\'ia, Universidad de Valpara\'iso, Gran Breta\~na 1111,
Valpara\'iso, Chile,}
\affiliation{Centro de Astrof\'isica de Valpara\'iso, Gran Breta\~na 1111, Playa Ancha,
\\ Valpara\'{\i}so, Chile.}

\date{\today}

\begin{abstract}

In this letter we study the semi holographic model which
corresponds to the radiative version of the model
proposed by Zhang et al. (Phys. Lett. \textbf{B} 694 (2010), 177) and revisited by C\'ardenas et al. (Mon. Not. Roy. Astron. Soc. 438 (2014), 3603).
This inclusion makes the model more realistic, so allows us to test it with current observational data and then answer if the inconsistency reported by C\'ardenas et al. is relaxed. 

\end{abstract}



\maketitle


At the moment we may say that the physical nature of the dark
sector of the universe still remains as a mystery in spite of the
efforts to consistently explain the observations coming from
supernovae of type Ia (SNIa)\cite{supernova}, large scale
structure (LSS)\cite{02}, cosmic microwave background
(CMB)\cite{03}, the integrated Sachs--Wolfe effect
(ISW)\cite{isw}, baryonic acoustic oscillations
(BAO)\cite{Eisenstein:2005su} and gravitational lensing
\cite{weakl}. One of the preferred cosmological model is the
so-called $\lambda$-cold-dark-matter ($\lambda CDM$). Although it
fits quite well most of the observational data, it suffers from
two main problems, namely: the low value of the vacuum energy
(about 120 magnitude orders below the quantum field theory
estimation)\cite{vacenergy} and the so-called coincidence
problem\cite{coincidence,dC08}. In order to solve these problems
it has been evoked to quintessence kind of models, where the
cosmological constant is substituted for a slowly-varying,
spatially inhomogeneous component with a negative equation of
state\cite{CDS98}. In this context, it is assumed a dynamical
cosmological constant, that leads to a dynamical dark energy
model, where a scalar field plays a central role (e.g., K-essence
\cite{3}, tachyon fields \cite{4}, etc.). Also, due to both dark
components are characterized through their gravitational effects,
it is natural to consider unified models of the cosmological
substratum in which one single component plays the role of DM and
DE simultaneously. Examples of this type of models are the
Chaplygin gas \cite{5,sergiojose,dC13-01,dC13-02}, and
bulk-viscous models \cite{bv}.

Among the scenarios that are based on a dynamical cosmological
constant we distinguish the holographic models\cite{holog,Julio}.
These models are achieved from the context of fundamental
principle of quantum gravitational physics, so called holographic
principle. This principle arose due to development in the study of
black hole physics together with superstring theories.
Specifically, it is derived with the help of entropy-area relation
of thermodynamics of black hole horizons in general relativity,
which is also known as the Bekenstein-Hawking entropy bound, i.e.,
$S \simeq \mu^2 L^2$, where $S$ is the maximum entropy of the
system of length $L$ and $\mu=1/\sqrt{8\pi\,G}$ is the reduced
Planck mass. Using this idea Cohen \cite{cohen} suggested a
relation between the short distance (ultraviolet, UV) cutoff and
the long distance (infrared, IR) cutoff which, after identifying
infrared with the Hubble radius $H^{-1}$, resulted in a DE density
very close to the observed critical energy density. After this, Li
\cite{li} studied the use of both the particle and event horizons
as the IR cutoff length. He found that apparently only a future
event horizon cutoff can give a viable DE model. More recently, it
was proposed a new cutoff scale, given by the Ricci scalar
curvature \cite{022}, resulting in the so-called holographic Ricci
DE models\cite{Winfried01,Winfried02}.

Apart of the models described above, it was proposed a model in
which the dark energy sector obeys strictly the holographic
principle\cite{Zhang:2010iz}. In this scenario it was found that
stable solutions exist that ameliorate the coincidence problem, but
out of an accelerated expansion\cite{Li:2012vw} why add radiation.  
In a preliminary study of this model \cite{Cardenas:2013ela} we showed
that is not necessary to add explicitly a cosmological constant to
get an accelerated expansion, and also we showed that this model can
fit well the supernovae data. However, because we were interested in
the low redshift transition from a decelerated phase to the
accelerated one, we only tested the performance of the data to fit
low redshift data. In this letter we intend to study this sort of
model, but where the radiation component is taken into account.
After of doing an analytical study we proceed to check the model
with recent observational data.




After reheating has occurred at the end of the inflationary period,
the universe evolves homogeneously and isotropically in an
adiabatical way. In a appropriated comoving volume $V$, expressed
by the physical volume $V=\frac{4}{3}\pi a^3$ where $a$ denotes
the scale factor, the first law of thermodynamics for a spatially
flat space reads
\begin{equation} dU + pdV = TdS, \label{1st}
\end{equation}
where $U=\frac{4}{3}\pi\rho a^3$ represents the energy in this
volume with $\rho$ the energy density, $T$ denotes the
temperature, $S$ represents the entropy and $p$ the pressure of
the related fluid.

Following Ref. \cite{bak,cai}, it is possible to write the entropy
$S$ associated to the apparent horizon in a comoving volume
as \cite{Zhang:2010iz}
   \be
   S_c=\frac{8\pi^2 \mu^2}{H^2} \frac{a^3}{H^{-3}}={8\pi^2
   \mu^2}Ha^3,
   \label{Sc}
   \en
where $\ds H \equiv \frac{\dot{a}}{a}$ is the Hubble parameter.
The semi holographic model emerges by assuming that the dark
energy component satisfies a similar relationship
    \be
    S_{de}={8\pi^2
   \mu^2}Ha^3.
   \label{detro}
    \en
Since we have assumed that the expansion of the universe evolves
adiabatically, the dark matter entropy contribution reads
    \be
    S_{dm}=C-S_{de},
    \label{dmtro}
    \en
where $C$ is a constant representing the total entropy of the
comoving volume.

Friedmann equation gives us an expression which relate the Hubble
parameter, $H$, with the corresponding energy densities (or
equivalently, the entropies via expressions (\ref{Sc}) and
(\ref{detro})), as
 \be
   H^2=\frac{1}{3\mu^2}(\rho_{dm}+\rho_{de}+\rho_r),
   \label{fried}
   \en
where $\rho_{dm}$ denotes the energy density associated to dark
matter, $\rho_{de}$ denotes the energy density for the dark energy
component, and $\rho_r$ denotes the energy density related to the
radiation component. We believe that the inclusion in the model of
this latter component  is necessary in order to make the model
more realistic, and thus compares it with other models that have
been put forward in the literature.

On the other hand, the holographic principle requires that the
temperature is related to the Hubble function as \cite{cai}
   \be
   T=\frac{H}{2\pi}.
   \label{tem}
   \en
This latter expression is well known for de Sitter spaces.

We can get the dark energy evolution equation by combining
equations (\ref{tem}), (\ref{fried}), and (\ref{detro}), with the
first law, (\ref{1st})
   \be
   \frac{2}{3}\rho_{de} '=\rho_{dm}(1-\omega_{dm})-\rho_{de}(1+3\omega_{de})
   +\frac{2}{3}\rho_{r},
   \label{evlde}
   \en
where a prime means $'=d/d\log a$, $\omega_{dm}$ indicates the
equation of state (EoS) parameter associated to the dark matter,
and $\omega_{de}$ represents the EoS parameter related to the dark
energy. Using instead of (\ref{detro}) the complementary relation
(\ref{dmtro}) for dark matter we find that

\be
   \frac{2}{3}\rho_{dm} '=-\rho_{dm}(3+\omega_{dm})+\rho_{de}(-1+\omega_{de})
   -\frac{2}{3}\rho_{r}.
   \label{evldm}
   \en
Finally, as is usual, the evolution of the radiative
component is dictated by

\begin{equation}\label{evrad}
\rho_r'=-4\rho_r.
\end{equation}

Thus, we have written a set of basic equations, expressions
(\ref{evlde}), (\ref{evldm}) and (\ref{evrad}), which in the
following we want to analyze.

In order to perform this analysis, we write this system as ${\bf
u'}=M{\bf u}$ where ${\bf u}=(\rho_{de}, \rho_{dm}, \rho_r)$ is
the vector containing the different energy density components and
$M$ represents the following matrix

\begin{equation}\label{matrix}
M = \frac{3}{2}\left[
      \begin{array}{ccc}
        -1-3\omega_{de} & 1-\omega_{dm} & 2/3 \\
        -1+\omega_{de} & -3-\omega_{dm}  & -2/3\\
        0& 0& 8/3
      \end{array}
    \right].
\end{equation}

A standard procedure leads to analytic solutions. These solutions
are linear combinations of the roots, $r$, of quadratic equations
\begin{equation}\label{detM}
\det{( M-rI)}=0.
\end{equation}

Explicitly, we found that
\begin{eqnarray}\label{as1}
\nonumber \rho_{de}&=& \mathfrak{f}(z)\left\{\frac{a_1}{(1+z)^{\frac{3}{4}\eta}}+
\frac{a_2}{1+z} +\frac{a_3}{(1+z)^{1+\frac{3\kappa}{2}}}\right\}\\
\rho_{dm}&=&
\mathfrak{f}(z)\left\{\frac{b_1}{(1+z)^{\frac{3}{4}\eta}}+\frac{b_2}{1+z}
+\frac{b_3}{(1+z)^{1+\frac{3\kappa}{2}}}\right\},
\end{eqnarray}
where
\begin{eqnarray}
\nonumber
a_1&=&- 2\,\kappa \,(1-3\omega_{dm}) \rho_{r}^{(0)}=-b_1,\\ \nonumber
  a_2&=& (1 +\varphi) \left[\rho_{de}^{(0)}(\kappa +3 \,\omega_{de}-\omega_{dm}-2)-2
  \rho_{dm}^{(0)} (1-\omega_{dm})\right]+\rho_{r}^{(0)} \left[\kappa(1-3   \omega_{dm})
  -3 (\omega_{dm}+1) (\omega_{de}-\omega_{dm})\right], \\
\nonumber
a_3 &=& (1+\varphi) \left[\rho_{de}^{(0)}  \left(\kappa -3
\omega _{de}+\omega _{dm}+2\right)+2 \rho_{dm}^{(0)}
   \left(1-\omega _{dm}\right)\right]+\rho_{r}^{(0)} \left[\kappa(1-3
   \omega_{dm}) +3 (\omega_{dm}+1) (\omega_{de}-\omega_{dm}) \right], \\
\nonumber
b_2&=& (1+\varphi) \left[2 \rho_{de}^{(0)}  (1-\omega_{de})+\rho_{dm}^{(0)}
(\kappa -3 \omega_{de}+\omega_{dm}+2)\right]-\rho_{r}^{(0)}
   \left[\kappa  (1-3 \omega_{de})+(9 \omega_{de}-7) (\omega_{de}-\omega_{de})\right]\\\nonumber
 b_3&=&(1+\varphi) \left[-2 \rho_{de}^{(0)}  (1-\omega_{de})+\rho_{dm}^{(0)}
 (\kappa +3 \omega_{de}-\omega_{dm}-2)\right]-\rho_{r}^{(0)}
   \left[\kappa  (1-3\omega_{de})-(9 \omega_{de}-7) (\omega_{de}-\omega_{dm})\right] \\ \nonumber
  \kappa &=&
  \sqrt{(\omega_{de}-\omega_{dm})(-8+9\omega_{de}-\omega_{dm})},\\
  \eta&=&\kappa +3\omega_{de}+\omega_{dm},
\end{eqnarray}
and finally the function
\begin{equation}\label{fdez}
\mathfrak{f}(z)=
\frac{(1+z)^{4+\frac{3}{4}\eta}}{2\,\kappa\,(1+\varphi)}.
\end{equation}
Clearly, to get real energy densities solutions, it is needed to
impose the condition
\begin{equation}\label{cond1}
(\omega_{de}-\omega_{dm})(-8+9\omega_{de}-\omega_{dm})\geq 0.
\end{equation}
We also need to consider the stability of the solutions. The
corresponding critical points of the system are
\begin{equation}\label{as4}
  \rho_{de}^{(C)}=0,\qquad \rho_{dm}^{(C)}=0,\qquad \rho_{r}^{(C)}=0.
\end{equation}
Note that, from (\ref{as1}) and (\ref{fdez}) these critical points
occur at $z = -1$. However, as we shall see soon, they depend on
the initial conditions. We find that one of these solutions
becomes zero for $ z > 0$.

In order to make a stability analysis of the dynamical system, we
study the perturbations around of the critical points. This drive
us to a general expression $T=M \delta$, where
\begin{equation} T=\left(
                     \begin{array}{c}
                        \frac{d \delta \rho_{de}}{d s} \\
                        \,\\
                        \frac{d \delta \rho_{dm}}{d s} \\
                        \, \\
                        \frac{d \delta \rho_{r}}{d s} \\
                     \end{array}
                   \right), \qquad \delta =\left(
                     \begin{array}{c}
                       \delta \rho_{de} \\
                       \,\\
                       \delta \rho_{dm} \\
                       \,\\
                       \delta \rho_{r} \\
                     \end{array}
                   \right),\label{as5}\end{equation}
and $M$ is the matrix (\ref{matrix}) introduced previously.


The corresponding eigenvalues are given by
\begin{equation}\label{as7}
  \lambda_{1, 2}=-\frac{3}{4}\left(4+3\omega_{de}+\omega_{dm} \pm  \kappa\right),
  \quad \lambda_3=4.
\end{equation}
which shows that the system is stable if and only if the condition
$\kappa<4+3\omega_{de}+\omega_{dm}$ is satisfied ($\lambda_1<0$).


In order to enhance the insight about the evolution of this model,
we manipulate Eqs. (\ref{evlde}) and (\ref{evldm}) to put them in a typical ``adiabatic
form'' as $\dot{\rho}+3H(\rho + p)=0$, and write down an explicit
form for the effective EoS parameter for each contribution.

Using (\ref{evlde}), the dark energy density equation leads to an effective
pressure term
\begin{equation}\label{peffde}
p_{eff}^{de}=\rho_{de}\left[-\frac{1}{2}+\frac{3}{2}w_{de}-\frac{\rho_{dm}}{\rho_{de}}(1-w_{dm})-\frac{\rho_r}{3\rho_{de}}
\right],
\end{equation}
where through the analytical expressions (\ref{as1}), we can obtain a
closed analytical formula depending on the ``bare'' EoS parameters
and the actual values $\rho_{de}(0)$ and $\rho_{dm}(0)$. The same
can be done with the dark matter density equation (8), which leads
to the effective pressure
\begin{equation}\label{peffdm}
p_{eff}^{dm}=\rho_{dm}\left[\frac{1}{2}+\frac{1}{2}w_{dm}+\frac{\rho_{de}}{2
\rho_{dm}}(1-w_{de})+\frac{\rho_r}{3\rho_{dm}} \right].
\end{equation}
This expression coincides with Eq.(28) in reference
\cite{Zhang:2010iz} when we neglect radiation $\rho_r=0$.


Using the analytical solution presented above, we can test it
against the observations to constrain the $\omega_{dm}$ and
$\omega_{de}$ parameters. In this section we describe the $H(z)$,
supernova Ia (SNIa), baryon acoustic oscillations (BAO), and
cosmic microwave background (CMB) dataset,  and we describe the
related method to analyze them.

{\noindent \bf {The case of $H(z)$}}

We use $28$ points of the Hubble parameter measurements in
$0.07\le z \le2.3$ compiled by \cite{farooq}. The $\chi^2_{H}$ can
be written as
\begin{equation}
\chi_{H}^2 = \sum_{i=1}^{28} \frac{ \left( H(z_{i})
-H_{obs}(z_{i})\right)^2 }{ \sigma_{H_i}^{2} },
\end{equation}
where $H_{obs}(z_{i})$ is the observed Hubble parameter, $H(z)$ is
the theoretical value for the model, and $\sigma_{H_i}^{2}$ the
observational error.

{\noindent \bf {The SNIa dataset} }

We also use the Union 2 sample consisting in 557 SNIa points in
the redshift range $0.511<z<1.12$ \citep{amanullah_2010}. The SNIa
data give the distance modulus as a function of redshift
$\mu_{obs}(z)$. Theoretically the distance modulus is a function
of the cosmology through the luminosity distance
\begin{equation}\label{dlzf}
d_L(z)=(1+z)\frac{c}{H_0}\int_0^z \frac{dz'}{E(z')},
\end{equation}
valid for a flat universe with $E(z)=H(z)/H_0$. Explicitly the
theoretical value is computed by $\mu(z)=
5\log_{10}[d_L(z)/\texttt{Mpc}]+25$. We fit the SNIa with the
cosmological model by minimizing the $\chi^2$ value defined by
\begin{equation}
\chi^2=\sum_{i=1}^{580}\frac{[\mu(z_i)-\mu_{obs}(z_i)]^2}{\sigma_{\mu
i}^2}.
\end{equation}

{\noindent \bf {Baryon Acoustic Oscillations}}

The BAO measurements considered in our analysis are obtained from
the WiggleZ experiment \citep{2011MNRAS.tmp.1598B}, the SDSS DR7
BAO distance measurements \citep{2010MNRAS.401.2148P}, and 6dFGS
BAO data \citep{2011MNRAS.416.3017B}.

The $\chi^2$ for the WiggleZ BAO data is given by
\begin{equation}
\chi^2_{\scriptscriptstyle WiggleZ} = (\bar{A}_{obs}-\bar{A}_{th})
C_{\scriptscriptstyle WiggleZ}^{-1}
(\bar{A}_{obs}-\bar{A}_{th})^T,
\end{equation}
where the data vector is $\bar{A}_{obs} = (0.474,0.442,0.424)$ for
the effective redshift $z=0.44,0.6$ and 0.73. The corresponding
theoretical value $\bar{A}_{th}$ denotes the acoustic parameter
$A(z)$ introduced by \citet{2005ApJ...633..560E}:
\begin{equation}
A(z) = \frac{D_V(z)\sqrt{\Omega_{m}H_0^2}}{cz},
\end{equation}
and the distance scale $D_V$ is defined as
\begin{equation}
D_V(z)=\frac{1}{H_0}\left[(1+z)^2D_A(z)^2\frac{cz}{E(z)}\right]^{1/3},
\end{equation}
where $D_A(z)$ is the Hubble-free angular diameter distance which
relates to the Hubble-free luminosity distance through
$D_A(z)=D_L(z)/(1+z)^2$. The inverse covariance
$C_{\scriptscriptstyle WiggleZ}^{-1}$ is given by
\begin{equation}
C_{\scriptscriptstyle WiggleZ}^{-1} = \left(
\begin{array}{ccc}
1040.3 & -807.5 & 336.8\\
-807.5 & 3720.3 & -1551.9\\
336.8 & -1551.9 & 2914.9
\end{array}\right).
\end{equation}

Similarly, for the SDSS DR7 BAO distance measurements, the
$\chi^2$ can be expressed as \citep{2010MNRAS.401.2148P}
\begin{equation}
\chi^2_{\scriptscriptstyle SDSS} =
(\bar{d}_{obs}-\bar{d}_{th})C_{\scriptscriptstyle
SDSS}^{-1}(\bar{d}_{obs}-\bar{d}_{th})^T,
\end{equation}
where $\bar{d}_{obs} = (0.1905,0.1097)$ is the datapoints at
$z=0.2$ and $0.35$. $\bar{d}_{th}$ denotes the distance ratio
\begin{equation}
d_z = \frac{r_s(z_d)}{D_V(z)}.
\end{equation}
Here, $r_s(z)$ is the comoving sound horizon,
\begin{equation}
 r_s(z) = c \int_z^\infty \frac{c_s(z')}{H(z')}dz',
 \end{equation}
where the sound speed $c_s(z) = 1/\sqrt{3(1+\bar{R_b}/(1+z)}$,
with $\bar{R_b} = 31500 \Omega_{b}h^2(T_{CMB}/2.7\rm{K})^{-4}$ and
$T_{CMB}$ = 2.726K.

The redshift $z_d$ at the baryon drag epoch is fitted with the
formula proposed by \citet{1998ApJ...496..605E},
\begin{equation}
z_d =
\frac{1291(\Omega_{m}h^2)^{0.251}}{1+0.659(\Omega_{m}h^2)^{0.828}}[1+b_1(\Omega_b
h^2)^{b_2}],
\end{equation}
where
\begin{eqnarray}
&b_1 = 0.313(\Omega_{m}h^2)^{-0.419}[1+0.607(\Omega_{m}h^2)^{0.674}], \\
&b_2 = 0.238(\Omega_{m}h^2)^{0.223}.
\end{eqnarray}

$C_{\scriptscriptstyle SDSS}^{-1}$ in Eq. (12) is the inverse
covariance matrix for the SDSS data set given by
\begin{equation}
C_{\scriptscriptstyle SDSS}^{-1} = \left(
\begin{array}{cc}
30124 & -17227\\
-17227 & 86977
\end{array}\right).
\end{equation}

For the 6dFGS BAO data \citep{2011MNRAS.416.3017B}, there is only
one data point at $z=0.106$, the $\chi^2$ is easy to compute:
\begin{equation}
\chi^2_{\scriptscriptstyle 6dFGS} =
\left(\frac{d_z-0.336}{0.015}\right)^2.
\end{equation}

The total $\chi^2$ for all the BAO data sets thus can be written
as
\begin{equation}
\chi^2_{BAO} = \chi^2_{\scriptscriptstyle WiggleZ} +
\chi^2_{\scriptscriptstyle SDSS} + \chi^2_{\scriptscriptstyle
6dFGS}.
\end{equation}

{\noindent \bf{The Cosmic Microwave Background data}}

We also include CMB information by using the WMAP 9-yr data
\citep{cmb2} to probe the expansion history up to the last
scattering surface. The $\chi^2$ for the CMB data is constructed
as
\begin{equation}\label{cmbchi}
 \chi^2_{CMB} = X^TC_{CMB}^{-1}X,
\end{equation}
where
\begin{equation}
 X =\left(
 \begin{array}{c}
 l_A - 302.40 \\
 R - 1.7246 \\
 z_* - 1090.88
\end{array}\right).
\end{equation}
Here $l_A$ is the ``acoustic scale'' defined as
\begin{equation}
l_A = \frac{\pi d_L(z_*)}{(1+z)r_s(z_*)},
\end{equation}
where $d_L(z)=D_L(z)/H_0$ and the redshift of decoupling $z_*$ is
given by \citep{husugi},
\begin{equation}
z_* = 1048[1+0.00124(\Omega_b h^2)^{-0.738}]
[1+g_1(\Omega_{m}h^2)^{g_2}],
\end{equation}
\begin{equation}
g_1 = \frac{0.0783(\Omega_b h^2)^{-0.238}}{1+39.5(\Omega_b
h^2)^{0.763}},
 g_2 = \frac{0.560}{1+21.1(\Omega_b h^2)^{1.81}},
\end{equation}
The ``shift parameter'' $R$ defined as \citep{BET97}
\begin{equation}
R = \frac{\sqrt{\Omega_{m}}}{c(1+z_*)} D_L(z).
\end{equation}
$C_{CMB}^{-1}$ in Eq. (\ref{cmbchi}) is the inverse covariance
matrix,
\begin{equation}
C_{CMB}^{-1} = \left(
\begin{array}{ccc}
3.182 & 18.253 & -1.429\\
18.253 & 11887.879 & -193.808\\
-1.429 & -193.808 & 4.556
\end{array}\right).
\end{equation}

\begin{figure}
  \begin{center}
    \includegraphics[width=80mm]{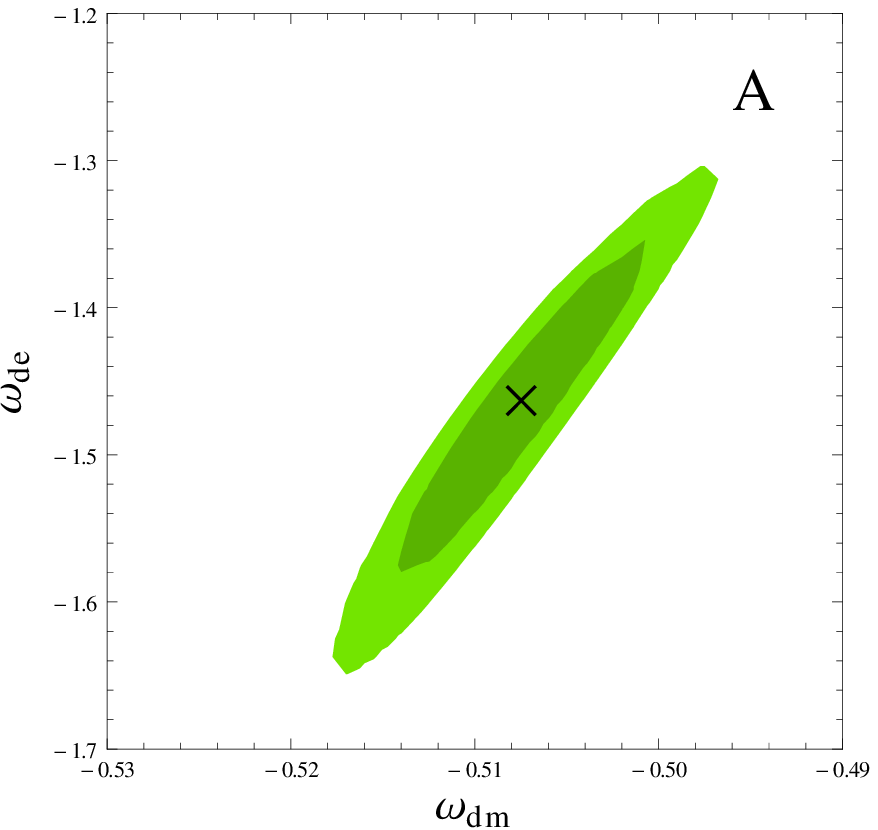}
    \includegraphics[width=80mm]{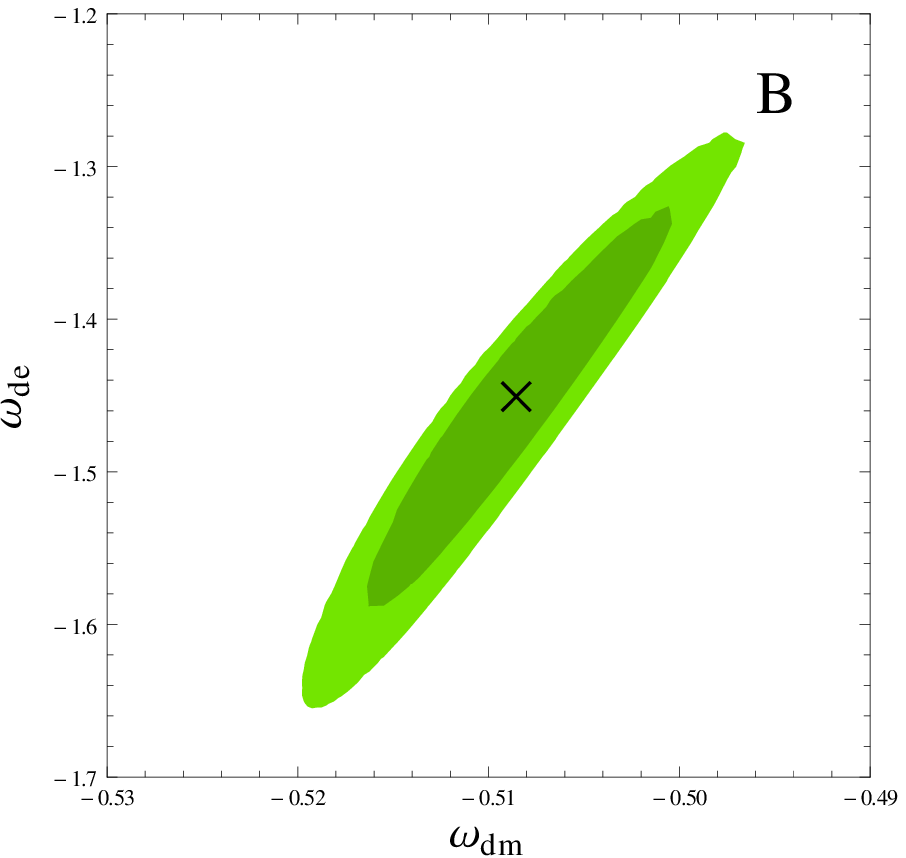}
  \end{center}
  \caption{Confidence contour at $1 \sigma$ and $2 \sigma$
  for case A (left panel) and case B (right panel) using all data described in Table 1. The cross indicates
  the best fit.}
  \label{f1}
\end{figure}

Assuming a flat geometry, we consider two cases to obtain
the best fit to the EoS of dark matter and dark energy. First in case A, we
assume the prior $\Omega_{dm}=0.235$ \citep{Hinshaw:2012} and we consider as
free parameters $h$, $\omega_{dm}$, and $\omega_{de}$. The
constraints for case A are given in Table \ref{tab:table01} using
different data sets. The best fit with $\chi^2_{min}=664.56$ derived
from the joint analysis including all data sets is $h=0.61$,
$\omega_{de}=-0.5$, and $\omega_{dm}=-1.46$  with confidence contours
at $68.27\%$ and $95.45\%$ shown in the first panel of Figure \ref{f1}.
Furthermore, we consider a second run -- which we called
case B -- in which $\Omega_{dm}$ is also considered a free
parameter, whose results are displayed also in Table \ref{tab:table01}.
Although the best fit parameters point out towards a
negative value for $\Omega_{dm}$ using $H(z)$ and $H(z)+$SNIa,
this behavior does not shows up once we include BAO data.
The best fit with $\chi^2_{min}=664.94$ derived from the joint
analysis including all data sets is $\Omega_{dm}=0.22$, $h=0.62$,
$\omega_{de}=-0.5$, and $\omega_{dm}=-1.45$. Notice the constraints
are very similar in both cases.

\begin{table}
\caption{The best fit values for the free
parameters using several data sets in a flat universe. We also
show the $\chi^2_{min}$ of the fit divided by the effective
degrees of freedom.} \centering
\begin{tabular}{lccccc}
Case&$\chi^2_{min}/d.o.f.$ & $\Omega_{dm}$ &$h$ & $\omega_{dm}$ & $\omega_{de}$ \\
\hline
\multicolumn{6}{|c|}{\small{$H(z)$}}\\
\hline
A&$15.53/28$ & Fixed & $0.72 \pm 0.05$& $-0.57 \pm 0.12$ &$ -1.38 \pm 0.67$\\
B&$15.85/28$ & $-0.03 \pm 0.51 $& $0.81 \pm 0.01$ & $-0.57 \pm 0.04$ &$ -1.65 \pm 0.08$\\
\hline
\multicolumn{6}{|c|}{$H(z)$+SNIa}\\
\hline
A&$559.90/585$ & Fixed &$0.68 \pm 0.01$& $-0.48 \pm 0.02$ & $-0.94\pm0.07$\\
B&$558.82/585$ & $-0.25$& $0.68$& $-0.33$ & $-0.68$\\
\hline
\multicolumn{6}{|c|}{$H(z)$+SNIa+BAO}\\
\hline
A&$594.01/591$ & Fixed & $0.63$& $-0.41$ & $-1.14$\\
B&$589.81/591$ & $0.27$ &$0.62$& $-0.41$ & $-1.17$\\
\hline
\multicolumn{6}{|c|}{$H(z)$+SNIa+BAO+CMB}\\
\hline
A&$664.56/594$ & Fixed &$0.61$ & $-0.5$ & $-1.46$\\
B&$663.94/594$ & $0.22$ & $0.61$ & $-0.5$ & $-1.45$\\
\hline
\label{tab:table01}
\end{tabular}
\end{table}

\begin{figure}
  \begin{center}
    \includegraphics[width=80mm]{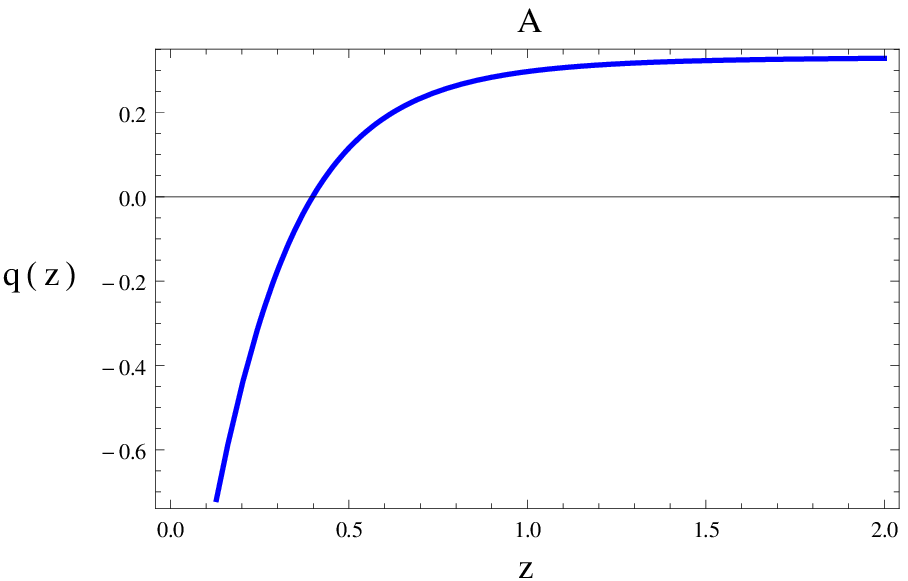}
    \includegraphics[width=80mm]{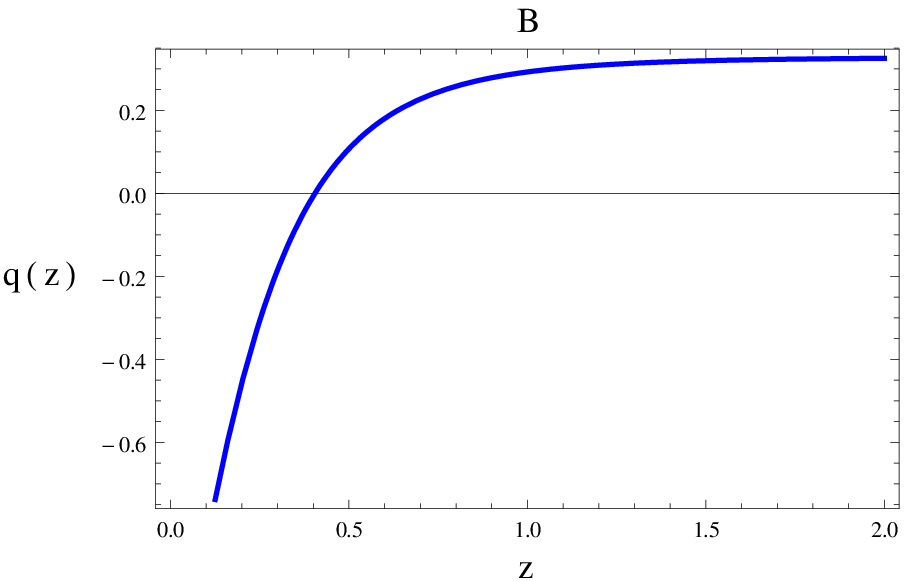}
  \end{center}
  \caption{The deceleration parameter reconstructed from the best fit for case A (left panel) and case B (right panel) using all data.}
  \label{f2}
\end{figure}

This means we can fit the observational data without using an
explicit cosmological constant, as was stressed in
\cite{Zhang:2010iz}, \cite{Cardenas:2013ela}. Actually,
reconstructing the deceleration parameter as a function of
redshift $q(z)$ (see Fig. \ref{f2}) we conclude that this model
describes an accelerated expansion without a cosmological constant.
The transition of a decelerated phase to an accelerated phase
occurs at $z\sim 0.4$.

\begin{figure}
  \begin{center}
    \includegraphics[width=75mm]{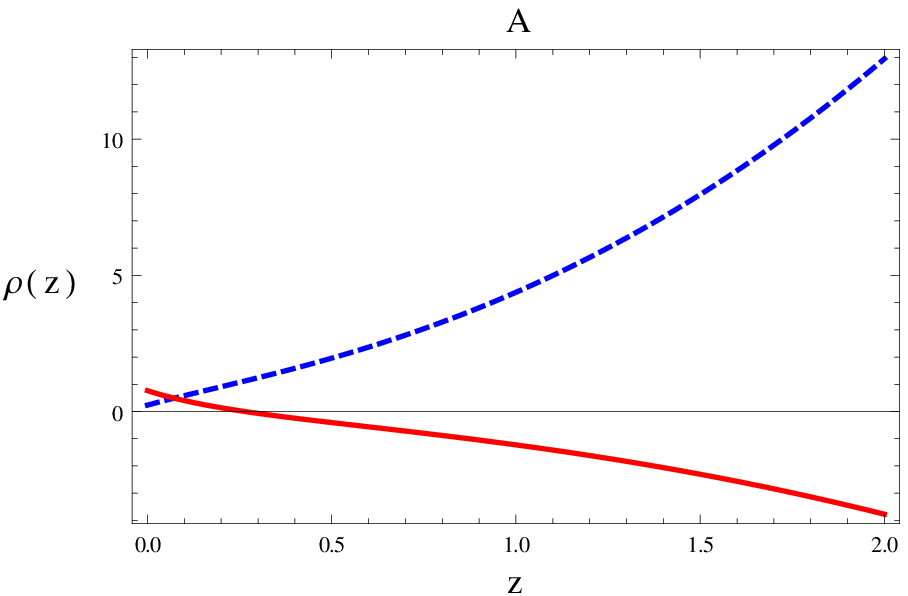}
     \includegraphics[width=75mm]{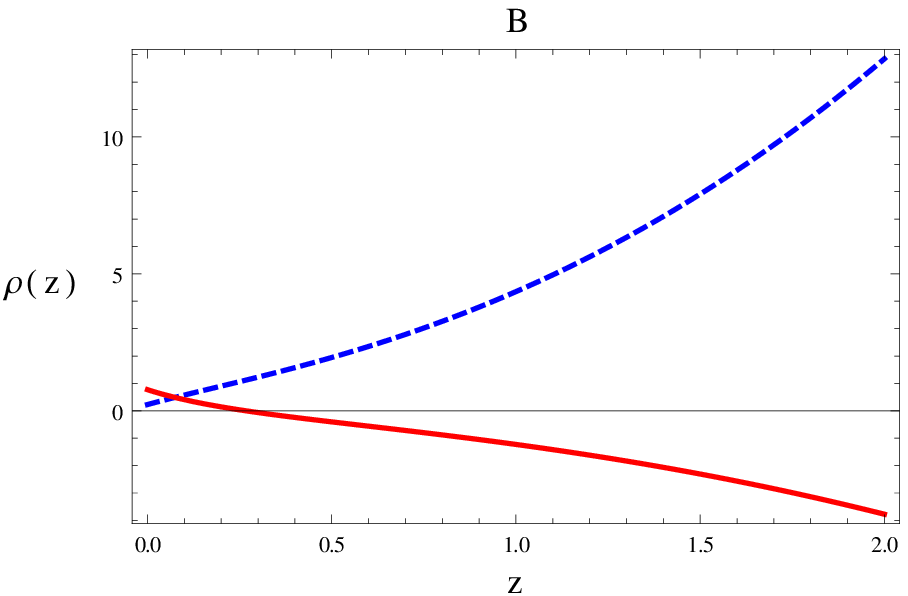}
  \end{center}
  \caption{Analytic solutions (\ref{as1}) of the energy
  densities, $\rho_{de}$ (red solid line) and $\rho_{dm}$ (blue dashed line) as a function of redshift, using
  the best fit values for the case A (left panel) and case B (right panel) using all data. Notice
  how $\rho_{de}$ falls to negative values near $z\simeq 0.25$.}
  \label{f3}
\end{figure}


However, as we have already found in the previous work
\cite{Cardenas:2013ela},  it seems that this model presents
a fatal failure since there is no chance to satisfied the weak
energy condition, because the energy density related to dark energy
becomes negative after a fraction of the redshift. In fact, this
can be viewed from the explicit solution that we have found,
expressed by equation (\ref{as1}). Based on the best fit values
found above, we have plotted the energy densities as a function of
redshift in Figure \ref{f3}, where it is clear that although both
energy densities are positive at $z=0$, i.e today, one of them (in
this case $\rho_{de}$) falls below zero near $z\simeq 0.25$.

It is interesting to note that this result is not
exceptional of this work. Actually in \cite{Cardenas:2014jya}  it
was shown evidence that relates the tension between low redshift
data and those from CMB, with a preference for evolving DE models
showing a decreasing DE density evolution with increasing redshift.
This results is also consistent with the recent BAO measurement of
BOSS DR11 \cite{Delubac:2014aqe}, which shows the data prefers a
decreasing DE density with increasing redshift. Although this
situation looks quite strange since never has been measured a
negative energy density, and it is possibly that this case
represents an unphysical situation in the sense that it may
violate the second law of thermodynamics \cite{2ndLaw} (although it
can be violated \cite{FT}), it is still exist the
possibility that a negative energy density studied in a background
of a FLRW metric results interesting \cite{rhonegative}. This could
be carried out since the energy conditions in relativity need only
be satisfied on a global scale, or on an average measure
\cite{Ford}. It might be interesting to address these issues deeply,
in the sense that these energy forms may describe cosmologically
interesting scenarios.

In this paper we have reviewed the semi-holographic model presented
by Zhang, Li and Noh \cite{Zhang:2010iz} plus radiation. From this, we have found
the analytical solutions to the dynamical system generated by this
model, Eqs. (\ref{as1}), finding accelerated expansion without a cosmological constant. Also, we
have studied the stability of the system, which drive us to impose
constraints on the EoS parameters, $\omega_{dm}$ and $\omega_{de}$,
Eq. (\ref{cond1}). Based on this analysis, we
determine the allowed region in parameter space to get stable
solutions, which does not agree with those mentioned in
\cite{Zhang:2010iz} (see eqs. (23) and (24) in this reference). The
analytical solution of the model was also tested against the latest
available observational data from Supernovas Ia, which is a set of
$580$ points for the module distance. 

\begin{acknowledgments}
This work was funded by the Comisi{\'o}n Nacional de Investigaci{\'o}n Cient{\'i}fica y Tecnol{\'o}gica through FONDECYT Grants No 1110230 (SdC, VHC)
and  No 11130695 (JRV). 
JM acknowledges ESO - Comit\'e Mixto, VHC
acknowledges also the financial support from DIUV project No.
13/2009.
\end{acknowledgments}


\begin{thebibliography}{99}


\bibitem{supernova}A. G. Riess et al., Astron. J. 116, 1009
(1998)[astro-ph/9805201 ]; S. J. Perlmutter et al., Astrophys. J.
517, 565(1999); A. G. Riess et al., Astrophys. J. 607, 665(2004).

\bibitem{02}M. Tegmark et al. [SDSS Collaboration], Phys. Rev. D 69, 103501
(2004); K. Abazajian et al. [SDSS Collaboration], Astron. J. 128,
502 (2004); K. Abazajian et al. [SDSS Collaboration], Astron. J.
129, 1755 (2005).

\bibitem{03} H. V. Peiris et al., Astrophys. J. Suppl. 148 (2003) 213 [astro-ph/0302225]; C. L.
Bennett et al., Astrophys. J. Suppl. 148  1 (2003); D. N. Spergel et
al., Astrophys. J. Suppl. 148  175 (2003).

\bibitem{isw}
S. Boughn and R. Chrittenden, Nature (London) \textbf{427}, 45
(2004); P. Vielva, E. Mart\'{\i}nez--Gonz\'{a}lez, and M. Tucci,
Mon. Not. R. Astron. Soc. \textbf{365}, 891 (2006).

\bibitem{Eisenstein:2005su}
  D.~J.~Eisenstein {\it et al.}  [SDSS Collaboration],
  Astrophys.\ J.\  {\bf 633}, 560 (2005)
  [astro-ph/0501171].

\bibitem{weakl}
C.R. Contaldi, H. Hoekstra, and A. Lewis, Phys. Rev. Lett.
\textbf{90}, 221303 (2003).

\bibitem{vacenergy} S. Weinberg, Rev. Mod. Phys. 61, 1 (1989);
 E.J. Copeland, M. Sami, S. Tsujikawa, Int. J. Mod. Phys. D 15, 1753 (2006).

\bibitem{coincidence}I. Zlatev, L. Wang and P. J. Steinhardt, Phys. Rev. Lett. {\bf 82}  896 (1999).

\bibitem{dC08}S. del Campo, R. Herrera and D. Pavon, Phys. Rev. D {\bf 78}, 021302(RC) (2008).

\bibitem{CDS98} R.R. Caldwell, R. Dave and P.J. Steinhardt, Phys. Rev. Lett. {\bf 80} 1582 (1998).


\bibitem{3}T. Chiba, T. Okabe, M. Yamaguchi, Phys. Rev. D 62, 023511 (2000);
C. Armend´ariz-Pic´on, V. Mukhanov, P.J. Steinhardt, Phys. Rev.
Lett. 85, 4438 (2000).

\bibitem{4} A. Sen, J. High Energy Phys. 10, 008 (1999); E.A. Bergshoeff, M.
de Roo, T.C. de Wit, E. Eyras, S. Panda, J. High Energy Phys. 05,
009 (2000).


\bibitem{5}A. Kamenshchik, U. Moschella, V. Pasquier, Phys. Lett. B 511, 265
(2001).

\bibitem{sergiojose}
S. del Campo and J. R. Villanueva,
Int. J. Mod. Phys. D \textbf{18},  2007-2022  (2009).

\bibitem{dC13-01}S. del Campo, Jour. Cosm. Astropart. Phys. (JCAP) {\bf 1311} 004
(2013).

\bibitem{dC13-02}S. del Campo, C. R. Fadragas, R. Herrera, C. Leiva, G. Leon and J.
Saavedra, Phys.Rev. D {\bf 88} 023532 (2013).

\bibitem{bv}
W. S. Hip\'{o}lito-Ricaldi, H.E.S. Velten and W. Zimdahl,
Phys. Rev. D  \textbf{82}, 063507 (2010).


\bibitem{holog}A. G. Cohen, D. B. Kaplan, and A. E. Nelson, Phys. Rev.
Lett. {\bf 82}, 4971 (1999); M. Li, Phys. Lett. B {\bf 603}, 1
(2004); S. D. H. Hsu, Phys. Lett. B {\bf 594}, 13 (2004).

\bibitem{Julio}S. del Campo, J. C. Fabris, R. Herrera and W. Zimdahl, Phys.Rev. D {\bf 83} 123006
(2011).


\bibitem{cohen}
A. G. Cohen, D.B. Kaplan and A.E. Nelson, Phys. Rev. Lett.
\textbf{82}, 4971 (1999).

\bibitem{li}
M. Li, Phys. Lett. B \textbf{603}, 1 (2004).


\bibitem{022} C. J. Feng, Phys. Lett. B {\bf 670}, 231 (2008);
~L. N. Granda and A. Oliveros, Phys. Lett. B {\bf 669}, 275 (2008).



\bibitem{Winfried01} S. del Campo, J. C. Fabris, R. Herrera and W. Zimdahl, Phys. Rev.
D {\bf 87}, 123002 (2013).

\bibitem{Winfried02}W. Zimdahl, J.C. Fabris, S. del Campo and R.
Herrera, Cosmology with Ricci-type dark energy. Conference:
C13-05-27.4, e-Print: arXiv:1403.1103[astro-ph.CO].


\bibitem{Zhang:2010iz}
  H.~Zhang, X.~-Z.~Li and H.~Noh,
  Phys.\ Lett.\ B {\bf 694}, 177 (2010)
  [arXiv:1010.1362 [gr-qc]].

\bibitem{Li:2012vw}
  H.~Li, H.~Zhang and Y.~Zhang,
  arXiv:1212.2360 [astro-ph.CO].

\bibitem{Cardenas:2013ela}
  V.~H.~C\'ardenas, J.~R.~Villanueva and J.~Maga\~na,
  Mon.\ Not.\ Roy.\ Astron.\ Soc.\  {\bf 438}, 3603 (2014)
  [arXiv:1306.6612 [astro-ph.CO]].


\bibitem{bak}
 D. Bak and S. J. Rey, Class. Quant. Grav. {\bf 17}, L83 (2000).

\bibitem{cai}
  R.~G.~Cai and S.~P.~Kim,
  JHEP {\bf 0502}, 050 (2005).

\bibitem {farooq} O. Farooq and B. Ratra, Ap. J. {\bf 766}, L7 (2013).

\bibitem{amanullah_2010}
R. {Amanullah}  {et~al.}, ApJ, {\bf 716}, 712 (2010).

\bibitem[{{Blake} {et~al}\mbox{.}(2011){Blake}, {Kazin}, {Beutler}, {Davis},
  {Parkinson}, {Brough}, {Colless}, {Contreras}, {Couch}, {Croom}, {Croton},
  {Drinkwater}, {Forster}, {Gilbank}, {Gladders}, {Glazebrook}, {Jelliffe},
  {Jurek}, {Li}, {Madore}, {Martin}, {Pimbblet}, {Poole}, {Pracy}, {Sharp},
  {Wisnioski}, {Woods}, {Wyder}, \& {Yee}}]{2011MNRAS.tmp.1598B}
{Blake} C. {et~al.}, 2011, MNRAS, 1598

\bibitem[{{Percival} {et~al}\mbox{.}(2010){Percival}, {Reid}, {Eisenstein},
  {Bahcall}, {Budavari}, {Frieman}, {Fukugita}, {Gunn}, {Ivezi{\'c}}, {Knapp},
  {Kron}, {Loveday}, {Lupton}, {McKay}, {Meiksin}, {Nichol}, {Pope},
  {Schlegel}, {Schneider}, {Spergel}, {Stoughton}, {Strauss}, {Szalay},
  {Tegmark}, {Vogeley}, {Weinberg}, {York}, \& {Zehavi}}]{2010MNRAS.401.2148P}
{Percival} W.~J. {et~al.}, 2010, MNRAS, 401, 2148

\bibitem[{{Beutler} {et~al}\mbox{.}(2011){Beutler}, {Blake}, {Colless},
  {Jones}, {Staveley-Smith}, {Campbell}, {Parker}, {Saunders}, \&
  {Watson}}]{2011MNRAS.416.3017B}
{Beutler} F. {et~al.}, 2011, MNRAS, 416, 3017

\bibitem[{{Eisenstein} {et~al}\mbox{.}(2005){Eisenstein}, {Zehavi}, {Hogg},
  {Scoccimarro}, {Blanton}, {Nichol}, {Scranton}, {Seo}, {Tegmark}, {Zheng},
  {Anderson}, {Annis}, {Bahcall}, {Brinkmann}, {Burles}, {Castander},
  {Connolly}, {Csabai}, {Doi}, {Fukugita}, {Frieman}, {Glazebrook}, {Gunn},
  {Hendry}, {Hennessy}, {Ivezi{\'c}}, {Kent}, {Knapp}, {Lin}, {Loh}, {Lupton},
  {Margon}, {McKay}, {Meiksin}, {Munn}, {Pope}, {Richmond}, {Schlegel},
  {Schneider}, {Shimasaku}, {Stoughton}, {Strauss}, {SubbaRao}, {Szalay},
  {Szapudi}, {Tucker}, {Yanny}, \& {York}}]{2005ApJ...633..560E}
{Eisenstein} D.~J. {et~al.}, 2005, ApJ, 633, 560

\bibitem[{{Eisenstein} \& {Hu}(1998)}]{1998ApJ...496..605E}
{Eisenstein} D.~J., {Hu} W., 1998, ApJ, 496, 605

\bibitem[{{Bond}, {Efstathiou} \& {Tegmark}(1997){Bond}, {Efstathiou}, \&
  {Tegmark}}]{BET97}
{Bond} J.~R., {Efstathiou} G., {Tegmark} M., 1997, MNRAS 291, L33

\bibitem[\protect\citeauthoryear{Hinshaw et al.}{2012}]{Hinshaw:2012}
G.~Hinshaw, {et~al}, 2013, ApJS, 208, 19.

\bibitem{Cardenas:2014jya}
  V.~H.~Cardenas,
  arXiv:1405.5116 [astro-ph.CO].

\bibitem{Delubac:2014aqe}
  T.~Delubac {\it et al.}  [BOSS Collaboration],
  arXiv:1404.1801 [astro-ph.CO].


\bibitem{2ndLaw} S. W. Hawking and G. F. R. Ellis, The large scale structure
of spacetime (University of Chicago Press, 1973).

\bibitem{FT}L. H. Ford and T. A. Roman, Phys. Rev. D {\bf 64}, 024023
(2001).

\bibitem{rhonegative}T. Hertog, G. T. Horowitz, K. Maeda, JHEP {\bf 0305}, 060 (2003);
R. J. Nemiroff, R. Joshi and B. R. Patla, An exposition on
Friedmann cosmology with negative energy densities,
arXiv:1402.4522[astro-ph.CO].


\bibitem{Ford} L. H. Ford and T. A. Roman, Phys. Rev. D {\bf 53},
1988 (1996); ibid, Phys. Rev. D {\bf 77}, 045018  (2008); M. J.
Pfenning, arXiv:gr-qc/9805037; M. Visser and C. Barcelo,
arXiv:gr-qc/0001099.

\bibitem[{{Hinshaw et al.}(2012)}]{cmb2}
  G.~Hinshaw, D.~Larson, E.~Komatsu, D.~N.~Spergel, C.~L.~Bennett, J.~Dunkley, M.~R.~Nolta and M.~Halpern {\it et al.},
[arXiv:1212.5226 [astro-ph.CO]].

\bibitem[{{Hu \& Sugiyama}(1996)}]{husugi}
{Hu} W., {Sugiyama} N., 1996, ApJ 471, 542

\end{thebibliography}
\end{document}